\documentclass[a4paper]{PoS}
\usepackage{cite}
\usepackage[utf8]{inputenc} 
\usepackage[mathscr]{euscript}
\usepackage{microtype}

\title{Effective-particle approach to bound states of quarks and gluons in QCD}

\ShortTitle{Effective-particle approach to bound states of quarks and gluons in QCD}

\author{{Mar\'ia G\'omez-Rocha}\thanks{Speaker, oral contribution}\\
        European Centre for Theoretical Studies in Nuclear Physics and Related Areas (ECT*)\\
        E-mail: \email{mariagomezrocha@gmail.com}}

\author{{Kamil Serafin}\thanks{Speaker, poster contribution}\\
        University of Warsaw - Faculty of Physics\\
        E-mail: \email{Kamil.Serafin@fuw.edu.pl}}

\abstract{
A general approach to the construction of bound states in quantum field theory,
called the renormalization group procedure for effective particles (RGPEP),
was applied recently to single heavy-flavor QCD in order to study its utility
beyond illustration of its general features. This heavy-flavor QCD is chosen
as the simplest available context in which the dynamics of quark and gluon bound
states can be studied with the required rigor using Minkowski-space Hamiltonian
operators in the Fock space, taking the advantage of asymptotic freedom.
The effective quarks and gluons differ from the point-like canonical ones
by having a finite size $s$. Their size plays the role of renormalization
group parameter. However, instead of integrating out high-energy degrees
of freedom, our RGPEP procedure is based on a transformation of the front-form
QCD Hamiltonian from its canonical form with counterterms to the renormalized,
scale-dependent operator that acts in the Fock space of effective quanta of
quark and gluon fields, keeping all degrees of freedom intact but accounting
for them in a transformed form. We discuss different behavior of effective
particles interacting at different energy scales, corresponding to different
size s. Namely, we cover phenomena ranging from asymptotic freedom at highest
energies down to the scales at which the formation of bound states occurs.
We briefly present recent applications of the RGPEP to quarks and gluons in QCD,
which have been developed using expansion in powers of the Fock-space Hamiltonian
running coupling. After observing that the QCD effective Hamiltonian satisfies
the requirement of producing asymptotic freedom, we derive the leading effective
interaction between quarks in heavy-flavor QCD. An effective confining effect is
derived as a result of assuming that the non-Abelian and non-perturbative dynamics
causes effective gluons to have mass. 
 }

\FullConference{XVII International Conference on Hadron Spectroscopy and Structure - Hadron2017\\
		25-29 September, 2017\\
		University of Salamanca, Salamanca, Spain}

\newcommand{\bal}{\begin{align}}
\newcommand{\eal}{\end{align}}
\newcommand{\beq}{\begin{eqnarray}}
\newcommand{\eeq}{\end{eqnarray}}
\newcommand{\nneeq}{\nonumber \end{eqnarray}}
\newcommand{\nnn}{\nonumber}

\newcommand{\es}{& = &}
\newcommand{\rs}{\, = \,}

\newcommand{\cM}{ {\cal M} }
\newcommand{\cH}{ {\cal H} }

\newcommand{\cU}{ {\cal U} }

\newcommand{\black}{\color[rgb]{0,0,0}}

\newcommand{\h}{ {1 \over 2} }

\newcommand{\ket}[1]{ {|{#1}\rangle} }
\newcommand{\bra}[1]{ {\langle{#1}|} }

\newcommand{\bmat}{\left[\begin{array}}
\newcommand{\emat}{\end{array}\right]}

\begin{document}

\section{ Introduction }

In spite of decades of research the bound-state equation in quantum
chromodynamics remains to be an exhausting problem without an exact
solution. The elements that build the Schr\"odinger-like equation
in QCD, $\hat H\ket{\psi}=E\ket{\psi}$, which gives hadron masses
and wave functions in the Fock space are full of complexities. It is
not straightforward how to determine the Hamiltonian whose eigenvalues
correspond to hadron masses starting from the QCD Lagrangian density
of the theory and using first principles. 
 
The main source of difficulties concerns the fact that in quantum
field theory one is forced to deal with an infinite number of degrees
of freedom. For instance, a heavy-quarkonium state -- the simplest
possible system one may consider in QCD -- has the structure of an
infinite series of Fock components 
\begin{eqnarray}
\ket{\psi}=\ket{QQ}+\ket{QQG}+\ket{QQGG}+\dots \ , 
\label{QCDstate}
\end{eqnarray}
and in principle, there is no limit on the number of particles allowed.

The renormalization group procedure for effective particles (RGPEP) was
formulated as a nonperturbative tool for constructing bound states in QFT.
The method stems from the similarity renormalization group (SRG) for
Hamiltonians~\cite{SRG1} (see also~\cite{Wilsonetal}) and introduces
the concept of effective particles. It provides a framework for the
description of the interaction of particles at different energy scales.
The main idea of the RGPEP is that it is possible to relate the canonical
Hamiltonian operator obtained from QCD with an effective one by means of
a similarity transformation. The Hamiltonian is written in a scale-dependent
operator basis which is such that, for a certain scale, the number of
nonnegligible Fock components in the description of hadrons is small.
When an infinite number of terms can be neglected in Eq.~(\ref{QCDstate}\black),
the bound-state problem is thus drastically simplified and one can attempt
to seek numerical solutions to the bound-state equation.

In this paper, we summarize the most recent results in the application
of the RGPEP to QCD~\cite{AF,QQbarRGPEP}. We start presenting the main
elements of the RGPEP method in the next section. The procedure is general
enough and can be applied to any theory. Afterwards, in
Section~\ref{AFsection}\black ~we consider QCD and show how the
property of asymptotic freedom arises in the formalism.
In Section~\ref{SecOutset}\black, we consider a theory of only one
heavy flavor in QCD and use asymptotic freedom to formulate the
bound-state equation for heavy quarkonium. We study the effective
potential obtained in the bound-state equation in Section~\ref{sectionResult}\black.
Section~\ref{Conclusions}\black  ~concludes the article.

\section{ RGPEP }

\subsection{ Initial Hamiltonian }

The starting point of our formulation of the bound-state equation is
the classical Lagrangian density of the theory. In this work, we are
concerned with the Lagrangian of QCD. The Noether theorem provides
the energy-momentum tensor $T^{\mu\nu}$ of the theory. We use the
front form (FF) of Hamiltonian dynamics~\cite{Dirac1949}, often called
light-front dynamics.\footnote{We adopt the notation and conventions
given in~\cite{Brodsky-Pauli-Pinsky}. The FF coordinates are defined
as $x^\mu=(x^+ = x^0 + x^3 , x^- = x^+ - x^3, \vec x^\perp)$, with
$x^\perp = (x^1,x^2)$. } 
The canonical FF Hamiltonian is obtained by integrating the $+-$ component
of the energy momentum tensor $T^{\mu\nu}$ over the quantization surface
$x^+=0$ in the gauge $A^+=0$, and replacing the classical fields by
quantum ones in\footnote{We use the usual Fourier decomposition,
\begin{eqnarray}
\hat A^\mu \es \sum_{\sigma c} \int [k] \left[ t^c \varepsilon^\mu_{k\sigma}
\hat a_{k\sigma c} e^{-ikx} + t^c \varepsilon^{\mu *}_{k\sigma}
\hat a^\dagger_{k\sigma c} e^{ikx}\right]_{x^+=0} \ , \quad
\hat\psi \ = \ \sum_{\sigma c } \int [k] 
  \left[ \chi_c u_{k\sigma} \hat b_{k\sigma c } e^{-ikx} + 
  \chi_c v_{k\sigma} \hat d^\dagger_{k\sigma c } e^{ikx}
  \right]_{x^+=0} \ .
  \nnn
\end{eqnarray}
}:
\begin{eqnarray}
H_{\rm can} \es P^- 
\ = \ \h \int dx^- d^2 x^\perp \, T^{+-}|_{x^+=0} \ .
\end{eqnarray}

The FF canonical Hamiltonian gives divergent integrals in perturbation
theory and needs to be regularized. We introduce smooth regulating functions
in interaction vertices, which depend on two regularization parameters:
$\Delta$, which regulates ultraviolet divergences related to big changes
of perpendicular momentum components $k^\perp$; and $\delta$, which regulates
the so-called small-$x$ divergences, related to $k^+$ component of the momentum.

\subsection{ The RGPEP equation }
The RGPEP introduces the concept of effective particles of size $s$.
The size $s$ is the renormalization group parameter. Creation and
annihilation operators labeled by $s$ create or annihilate effective
particles of size $s$:
\begin{eqnarray}
Q_0^\dagger \ket{0} = \ket{ Q_0 } \ , \quad 
Q_0\ket{0} = 0 \ , \quad
Q_s^\dagger \ket{0} = \ket{ Q_s } \ , \quad 
Q_s\ket{0} = 0 \ .
\end{eqnarray}
Effective particles of finite size $s$ and bare or canonical ones
of size 0 are related by a unitary transformation
\begin{eqnarray}
Q_s = \cU_s \, Q_0 \, \cU_s^\dagger \ .
\end{eqnarray}
For later convenience we define $t=s^4$, and express the
Hamiltonian in the new basis
\begin{eqnarray}
H_t(Q_t) = H_0 (Q_0) \ ,
\end{eqnarray}
where $H_t(Q_t)$ means that the Hamiltonian is expressed in terms
of effective operators $Q_t$ with effective coefficients standing
in front of them. The dependence of these coefficients with scale
$t$ is given by the equation
\begin{eqnarray}
\cH_t \es \left[ \left[ \cH_f , \cH_{Pt} \right] , \cH_t \right]
\ ,
\label{RGPEPeq}
\end{eqnarray}
where $\cH_t=H_t(Q_0)$, $\cH_f$ is the free part of $\cH_t$, and
$\cH_{Pt}$ is the same as $\cH_t$ but multiplied by the square of
total $+$-component of momentum entering the vertex.

Although Eq.~(\ref{RGPEPeq}\black) can be solved nonperturbatively,
at this stage and for the purposes of our current studies, we focus
on perturbative solutions to Eq.~(\ref{RGPEPeq}\black),
\begin{eqnarray}
H_t \es H_{t \, 0} + g H_{t\, 1} + g^2 H_{t\, 2}
+ g^3 H_{t\, 3} + g^4 H_{t\, 4} + \dots 
\end{eqnarray}

Solving Eq.~(\ref{RGPEPeq}\black) order by order yields exponentials
$\exp[-t(\cM_c^2-\cM_a^2)^2]$, which play the role of \textit{form factors}
appearing at interacting vertices, where $\cM_c$ and $\cM_a$ are invariant
masses of particles created and annihilated in a vertex~\cite{pRGPEP}. 
The effective Hamiltonian is determined by the initial condition that at
$t=0$ the Hamiltonian must equal the regularized canonical Hamiltonian
plus counterterms. The counterterms are determined by the condition that
every matrix element of the effective theory ($t>0$) is cutoff independent
when the ultraviolet regularization is removed.

The notion of effective particles can be also understood using the parameter
$\lambda=1/s$ which has dimension of mass. Namely, due to form factors,
effective particles of type $\lambda$ cannot change their relative kinetic
energy through a single effective interaction by more that about $\lambda$.

\section{ Asymptotic Freedom }
\label{AFsection}

The first check that was made in the application of the RGPEP to QCD was
its suitability to reproduce the property of asymptotic freedom~\cite{AF}
and the agreement of the result with the one obtained using another
generator~\cite{Glazek2000}. The feature of asymptotic freedom at short
distances could be checked  in terms of a family of renormalized effective
Hamiltonians using RGPEP. The structure of the three-gluon and quark-gluon
vertices can be extracted from third-order solutions to the RGPEP equation
for QCD. It was shown in Ref.~\cite{AF} that for a quantum Yang-Mills theory
the Hamiltonian running coupling evolves with the scale as
\begin{eqnarray}
g_\lambda \es
g_0 - { g_0^3 \over 48 \pi^2 }   N_c \,   11 \,\ln
{ \lambda \over \lambda_0} \ ,
\end{eqnarray}
which agrees with the known function obtained in~\cite{Politzer:1973fx,Gross:1973id}.
Whereas the running coupling described using Feynman diagrams evolves with
momentum scale, the Hamiltonian running coupling obtained within RGPEP evolves
with the parameter $\lambda$.  

We will make use of the property of asymptotic freedom to derive an effective
theory of heavy quarks in the next section. 

\section{ Heavy quarkonium problem }
\label{SecOutset}

The simplest bound system one can consider in QCD is heavy quarkonium.
To simplify the picture, in the following, we consider only one flavor
of heavy quarks and neglect light quarks. The eigenproblem may be simplified
drastically with the following choice of the renormalization group parameter,
\beq
m \gg \lambda  \gg \Lambda_{\rm QCD}
\label{scales}
\ ,
\eeq
where $m$ is the quark mass and $\Lambda_{\rm QCD}$ is the QCD scale.
On the one hand, $\lambda  \gg \Lambda_{\rm QCD}$ allows us to expand
the Hamiltonian in powers of $g_t=g$ and keep only the first few terms due to
asymptotic freedom. On the other hand, because $m \gg \lambda$, Fock sectors
with extra quark--antiquark pairs are strongly suppressed by RGPEP form factors
and we may neglect them. Sectors with gluons cannot be {\it a priori} neglected,
because gluons are massless and one may produce many of them without adding much
to the invariant mass of a system (contrary to the addition of heavy quarks).
This poses a problem because one cannot deal with infinitely many Fock sectors.
The solution we adopt is that we drop all the sectors with two or more gluons,
but make up for their absence by introducing a gluon mass ansatz in the sector
$Q_t\bar Q_tG_t$. The gluon mass ansatz is motivated by three observations.
First, it is the simplest term one can add. Second, an effective gluon mass
might arise as a nonperturbative effect when the reduction of the Fock space
is done exactly. This effect contains non-Abelian gluon--gluon interactions.
Third, the phenomenology of hadrons seems to exclude massless effective gluons. 

The heavy quarkonium bound-state problem with just two Fock sectors and gluon
mass ansatz is (see Ref.~\cite{QQbarRGPEP} for more details)  
\beq
\left[ \begin{array}{ll}
       H_f + \mu^2 \  &  g H_{t1}  \\  
       g H_{t1} \  & H_f + g^2 H_{t2}
       \end{array} \right]
\left[  \begin{array}{l} 
        | Q_t \bar Q_t G_t \rangle   \\
        | Q_t \bar Q_t \rangle 
        \end{array} \right] 
\es
E
\left[  \begin{array}{l} 
        | Q_t \bar Q_t G_t \rangle   \\
        | Q_t \bar Q_t \rangle 
        \end{array} \right] 
\  ,
\label{meson1}
\eeq
where $\mu^2$ is the gluon-mass-like operator acting within the $Q_t\bar Q_tG_t$
sector. An important assumption about $\mu^2$ is that it depends on the relative
motion of the gluon with respect to the quark-antiquark pair in the $Q_t\bar Q_tG_t$
sector. The need for this dependence is briefly explained in Sec.~\ref{sectionResult}\black.

Because we kept only terms order $1$, $g$ and $g^2$ in the approximate eigenvalue
problem Eq.~(\ref{meson1}\black), we may perturbatively eliminate the sector with
gluon reducing the eigenproblem to the lowest sector~\cite{Wilson1970}. Matrix
elements of the effective Hamiltonian obtained as a result of this elimination are
\beq
\bra{l} H_{t\,\rm{eff}} \ket{r}
\es
\bra{l} \left\{
  H_f
+ g^2 H_{t2}
+ \frac{1}{2} g H_{t1} \left[
    \frac{1}{E_l - H_f - \mu^2}
  + \frac{1}{E_r - H_f - \mu^2}
  \right] g H_{t1}
\right\} \ket{r}
\ ,
\label{Heff2nd}
\eeq
where $\ket{l}$ and $\ket{r}$ are both in $Q_t\bar Q_t$ sector
and $H_f\ket{l}=E_l\ket{l}$ and $H_f\ket{r}=E_r\ket{r}$.

\section{ Result: Coulomb and harmonic oscillator potentials }
\label{sectionResult}

\begin{figure}
    \centering
    \includegraphics[width=0.2\textwidth]{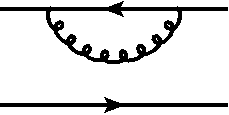}
    \hspace{1em}
    \includegraphics[width=0.2\textwidth]{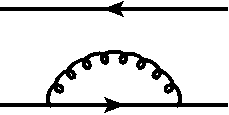}
    \hspace{1em}
    \includegraphics[width=0.2\textwidth]{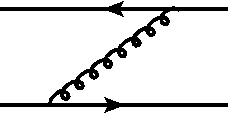}
    \hspace{1em}
    \includegraphics[width=0.2\textwidth]{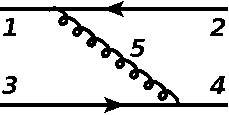}
    \caption{Self-interaction terms and gluon exchange terms in Eq.~(\ref{BS-eq-summary}).}
    \label{fig}
\end{figure}

As a result of elimination of the sector containing one gluon
we obtain a Hamiltonian acting in the lowest sector, $Q_t\bar Q_t$.
The FF eigenvalue equation is
\beq
H_{t\,\rm{eff}}|\psi_{Q\bar Q \,t}\rangle
\es
\frac{M^2 + P^{\perp 2}}{P^+} |\psi_{Q\bar Q \,t}\rangle
\ ,
\label{evestate}
\eeq
where $M$ is the quarkonium mass while $P^+$ and $P^\perp$ are longitudinal
and perpendicular momenta of the state
\beq
\ket{\psi_{Q\bar Q t}}
\es
\sum_{~~~~~ 24} \hspace{-15pt}\int  P^+ 
\frac{\delta_{c_{2}c_{4}} }{ \sqrt{3}}  \tilde\delta (P-k_2-k_4)
\, \psi_{t\,24}(\kappa^\perp_{24},x_2) 
b_{2t}^\dagger d_{4t}^\dagger \ket{0}
\ .
\label{wavefunc}
\eeq
$\delta_{c_{2}c_{4}}/\sqrt{3}$ is the color singlet wave function,
$\tilde\delta$ is a momentum conservation Dirac delta multiplied by
$16\pi^3$ and $\psi_{t\,24}(\kappa^\perp_{24},x_2)$ is a FF
wave function depending on spins (indicated by subscripts $24$,
cf. Figure~\ref{fig}\black), relative $\perp$-momentum $\kappa^\perp_{24}
= x_4 k_2^\perp - x_2 k_4^\perp$, and longitudinal momentum fraction
$x_i = k_i^+/P^+$ carried by the particle $i$. It is a property of the
front form that relative motion described by momenta $\kappa^\perp_{ij}$
and $x_i$ decouples from the absolute motion with momenta $P^+$ and $P^\perp$.
Therefore, the FF eigenvalues are in fact masses squared $M^2$.
Equation~(\ref{evestate}\black) is rewritten in terms of the FF wave
function,
\beq
&&
\left( 
  \frac{{\mathscr M}^2_{t} + \kappa_{13}^{\perp 2}}{ x_1 x_3 } 
- M^2 \right)\, \psi_{t\,13}(\kappa^\perp_{13},x_1) 
+ g^2 \int \frac{dx_2 d^2 \kappa_{24}^{\perp}}{ 2 (2\pi)^3 x_2 x_4} \, U_{t\,\rm{eff}}(13,24)
\, \psi_{t\,24}(\kappa^\perp_{24},x_{2})
\rs
0
\ .\hspace{1em}
\label{BS-eq-summary}
\eeq
${\mathscr M}^2_{t} = m^2 + \delta m_{t}^2$ is the quark mass
squared plus the second order effective quark self-interaction term,
$U_{t\,\rm{eff}}(13,24)$ is the effective interaction between quark
and antiquark coming from exchanging gluons and FF instantaneous
interactions (cf. Figure~\ref{fig}).

Equation~(\ref{BS-eq-summary}\black) with large relative momenta between
quarks suppressed by form factors due to $\lambda\ll m$ and with
small coupling constant due to $\lambda\gg\Lambda_{\rm QCD}$ may
be approximated by its nonrelativistic limit.
However, before looking for approximations we need to check if
all regularization dependence of Eq.~(\ref{BS-eq-summary}\black) is removed.
The problem is that there might appear small-x divergences when we
add an ansatz. For example, $\delta m^2_{t}$, which is zero in
the absence of gluon mass ansatz, is potentially divergent when small-x
regularization is lifted. Similar divergences are present also in the
exchange terms. To avoid any regularization dependence we have to assume
that $\mu^2$ vanishes in a proper way when $x_5\to0$. For example,
$\mu^2 \sim x_5^{\delta_\mu} \kappa_5^2$ when $x_5\to0$ (where $x_5$
and $\kappa_5^\perp$ are relative FF momenta of gluon with respect to
the $Q\bar Q$ pair) with $0<\delta_\mu<1/2$ is sufficient to guarantee that
the effective Hamiltonian is finite when
the small-x regularization is removed ($\delta\to0$). The validity of the gluon mass
ansatz may be checked by performing 4th order calculation of the effective
Hamiltonian for quarkonium. However, details of the gluon mass ansatz
turn out not to be important for the main result, Eq.~(\ref{schrodinger}\black).

To write the nonrelativistic approximation of Eq.~(\ref{BS-eq-summary}\black)
we need momentum variables which are more suitable than $\kappa^\perp$
and $x$. We define 
\beq
k^\perp_{ij} 
\rs \h \frac{ \kappa^\perp_{ij} }{ \sqrt{x_i x_j} } 
\ ,
\quad
k_{ij}^3 
\rs 
\frac{ m }{ \sqrt{ x_i x_j} } \left( x_i - \h \right)
\ ,
\label{kij}
\eeq
where $ij=13$ or $ij=24$, according to Fig.~\ref{fig}. 
We define also $M=2m+B$, where $B$ is binding energy, divide both
sides of Eq.~(\ref{BS-eq-summary}\black) by $4m$, and take the limit
$\vec k_{ij}/m\to0$. The result is
\beq
\left[
  { {\vec k}^{\, 2}_{13 } \over m} - B
  + \frac{\delta m_{t}^2 }{ m } 
\right]
\, \psi_{13}(\vec k_{13})
+ \int \frac{d^3 q}{(2\pi)^3} \, 
\left[V_{C,BF} + W ( \vec q\, )\right]
\, \psi_{24}(\vec k_{13}-\vec q)
\rs 0 \ ,
\eeq
where $V_{C,BF}$ stands for Coulomb potential with Breit-Fermi (BF)
spin-dependent interactions,
\beq
V_{C,BF}
\es
- \frac{4}{3}  
\frac{ 4 \pi \alpha }{ \vec q\,^2 } 
\left( 1 \ + \ BF \right)
e^{-16s^4(k_{13}^2-k_{24}^2)^2}
\ ,
\eeq
$4\pi\alpha=g^2$, $\vec q = \vec k_{13} - \vec k_{24}$, and
\beq
W(\vec q)
\es
-\frac{4}{3} 4\pi\alpha
\left(\frac{1}{q_z^{2}} - \frac{1}{\vec q\,^2}\right)
\frac{\mu^2}{\mu^2+\vec q\,^2} e^{-2tm^2 \frac{q^4}{q_z^2}}
\ .
\eeq
The exponential factor limits $q=|\vec q|$ to small values for
$\lambda\ll m$, hence $\mu^2/(\mu^2+q^2)\to1$ as long as $\mu\neq0$
and we can expand the wave function $\psi_{24}(\vec k_{13}-\vec q)$
at $\vec q=0$. The first term (independent of $\vec q$)
cancels with the mass term $\delta m_{t}^2/m = -\int d^3 q \, W(\vec q)
/(2\pi)^3$, while the terms linear in $\vec q$ vanish when integrated
over $\vec q$. Terms quadratic in $\vec q$ produce a correction
to Coulomb interaction proportional to $-\Delta_{\vec k}\psi=
r^2 \psi$, where $r$ is the distance between quark and
antiquark. The final result is
\beq
  { {\vec k}^{\, 2}_{13 } \over m} \, \psi_{13}(\vec k_{13})
+ \int \frac{d^3 q}{(2\pi)^3} \, V_{C,BF}
\, \psi_{24}(\vec k_{13}-\vec q)
+ \frac{1}{2} \frac{m}{2} \omega^2 r^2 \, \psi_{13}(\vec k_{13}\,)
\es
B \, \psi_{13}(\vec k_{13})
\label{schrodinger}
\ ,
\eeq
where the harmonic oscillator frequency is $\omega =
\sqrt{ \frac{\alpha}{18\sqrt{2\pi}} }\frac{\lambda^3}{m^2}$.

\section{ Conclusion }
\label{Conclusions}

The RGPEP is a tool suitable for the task of calculating bound
states in QCD. It passes the test of producing asymptotic freedom
in the running of the effective coupling constant and
in the second order calculation with a gluon mass ansatz
it produces a Schr\"odinger equation for
heavy quarks with Coulomb and harmonic oscillator potentials.
Harmonic oscillator potential in front form corresponds to
linear potential in the usual equal-time formulation~\cite{TrawinskiConfinement}.
Therefore, the result is expected to be a good first approximation
for calculations of hadron spectra. Moreover, the same harmonic
oscillator potential was found using a different version of
RGPEP~\cite{QQ1}, establishing a welcome degree of universality
of the harmonic oscillator potential result. Although this result
relies on the ansatz, the method presented here can be extended
to $g^4$ order and to include $Q_t\bar Q_tG_tG_t$ sector.
In such calculation, the gluon mass ansatz from $Q_t\bar Q_tG_t$
sector will be replaced by true QCD and the ansatz will be verified.
Another research goal is a nonperturbative determination
of running of effective quark and gluon masses with renormalization
scale $\lambda$.

\bibliographystyle{JHEP}
\bibliography{RGPEPrefs.bib}{}    

\end{document}